\documentclass[pra,twocolumn,showpacs,preprintnumbers,amsmath,amssymb]{revtex4}
\usepackage{amssymb}
\usepackage{amsfonts}
\usepackage{tipa}

\usepackage{epsfig,graphicx}
\usepackage{amstext}
\usepackage{amsmath}
\usepackage{graphicx}

\begin{document}


\title{Relations between entanglement, Bell-inequality violation and teleportation fidelity for the two-qubit X states}

\author{Ming-Liang Hu}
\email{mingliang0301@163.com}
\address{School of Science, Xi'an University of Posts and
               Telecommunications, Xi'an 710061, China}

\begin{abstract}
Based on the assumption that the receiver Bob can apply any unitary
transformation, Horodecki {\it et al.} [Phys. Lett. A {\bf 222}, 21
(1996)] proved that any mixed two spin-1/2 state which violates the
Bell-CHSH inequality is useful for teleportation. Here, we further
show that any X state which violates the Bell-CHSH inequality can
also be used for nonclassical teleportation even if Bob can only
perform the identity or the Pauli rotation operations. Moreover, we
showed that the maximal difference between the two average
fidelities achievable via Bob's arbitrary transformations and via
the sole identity or the Pauli rotation is $1/9$.
\end{abstract}


\maketitle

\section{Introduction}
Entanglement and Bell nonlocality are two aspects of quantum
correlations that have been the research interests since the early
days of quantum mechanics and still have not been completely
understood until now \cite{Nielsen,Bell,Clauser}. A basic trait of
the entangled state of a composite system is that it cannot be
written as products of states of each subsystem. As a physical
resource, entanglement is essential for various quantum information
processing (QIP) tasks \cite{Nielsen}. One of them is quantum
teleportation \cite{Bennett}, by which an unknown state can be
replicated at a distant location with the help of local operations
and classical communication. However, not all the states that are
entangled can be used for teleportation with average fidelity (see
sections below) better than that achievable via classical
communication alone \cite{Popescu,Horodecki1}, and the average
fidelity is even not a monotone function of the degree of
entanglement of the channel state. This seemingly counterintuitive
phenomenon has been noticed by a number of authors
\cite{Yeo1,Yeo2,Yeo3,Hu1,Hu2}. Particularly, there are situations in
which the channel states possess greater amount of entanglement with
however the average fidelity cannot exceed the classical limiting
value of $2/3$ \cite{Hu2}.

Bell nonlocality corresponds to another quantum correlation that
cannot be reproduced by any classical local-hidden variable models,
and this nonlocal behavior can be detected by the violation of
different Bell-type inequalities \cite{Nielsen}. The investigation
of Bell-nonlocality violation is significant not only for the
fundamental role it plays in better understanding of the subtle
aspects of quantum mechanics, but also because these violations are
crucial for some practical applications in quantum information
processing, such as to guarantee the safety of the
device-independent key distribution protocols in quantum
cryptography \cite{Acin,Gisin1}.

Due to their close relations, distinction between entanglement and
Bell-inequality violation has been studied extensively.
Particularly, it has been demonstrated that for the two-level
systems, the inseparability of a bipartite pure state corresponds to
the violation of the Bell-CHSH inequality, and vice versa
\cite{Horodecki1,Gisin2}. However, this is not the case for the
mixed states. As pointed out initially by Werner \cite{Werner},
there are bipartite mixed states who are entangled but do not
violate any Bell-type inequalities, thus one cannot distinguish
whether these correlations are produced by a classical local-hidden
variable model or not.

Moreover, when considering the issue of quantum teleportation
protocol, Popescu \cite{Popescu} noticed that there are bipartite
mixed states which do not violate any Bell-type inequalities, but
still can be used for teleportation with average fidelity larger
than the classical limiting value of $2/3$. It is then natural to
ask for the generic relations between entanglement, Bell-nonlocality
violation and quantum teleportation. In general, this problem is
rather complicated and difficult to answer. But for the special case
of the bipartite two-level systems, Horodecki {\it et al.} showed
that the question concerning the Bell-CHSH inequality violation and
the inseparability of the mixed states can be derived definitely
\cite{Horodecki1}. Based on the assumption that during the
teleportation process, the sender Alice uses only the Bell basis in
her measurement, while the receiver Bob is allowed to apply any
unitary transformation, Horodecki {\it et al.} demonstrated that any
two spin-1/2 state (pure or mixed) which violates the Bell-CHSH
inequality is useful for teleportation.

The statement of Horodecki {\it et al.} \cite{Horodecki1} relies
crucially on Bob's ability to perform any unitary transformation on
his qubit. But it is worthwhile to note that in real circumstances
the performance of certain unitary transformations, particularly for
solid-state construction of qubit systems, may be very difficult
\cite{Nielsen}, thus it is significant to consider the situation in
which Bob performs only some easy-realized transformations. For
instance, if Bob can only perform the identity (i.e., do nothing) or
the Pauli rotation operations, then what will happen to the
teleportation process? Can it still enable nonclassical fidelity
when the channel state violates the Bell-CHSH inequality
\cite{Clauser}. In fact, this is not the case as we will show in the
following text after introducing the definitions of the
corresponding average fidelities.

\section{Bell-inequality violation and teleportation via the X states}
In this paper, we will address the problem concerning possible
relations between entanglement, Bell-nonlocality violation and
quantum teleportation for the two-qubit X state $\rho$ \cite{Yu},
which has nonzero elements only along the main diagonal and
anti-diagonal. We assume that Alice uses the generalized Bell basis
$|\Psi^{0,3}\rangle=(|00\rangle\pm e^{-i\alpha}|11\rangle)/\sqrt{2}$
and $|\Psi^{1,2}\rangle=(|01\rangle\pm
e^{-i\beta}|10\rangle)/\sqrt{2}$ in her measurement, while Bob
performs only the identity or the Pauli rotation operation to his
qubit according to the classical information he received form Alice
\cite{Bennett}. Here the exponential terms $e^{-i\alpha}$ and
$e^{-i\beta}$ are related to the matrix elements of the X state via
$w=|w|e^{i\alpha}$ and $z=|z|e^{i\beta}$. The X state can be written
in the following form
\begin{equation}
 \rho=\left(\begin{array}{cccc}
  a   & 0   & 0  & w \\
  0   & b   & z  & 0 \\
  0   & z^* & c  & 0 \\
  w^* & 0   & 0  & d \\
 \end{array}\right).
\end{equation}

Such X states arise in a wide variety of physical situations and
include pure Bell states \cite{Nielsen} as well as the well-known
Werner-like mixed states \cite{Werner}. The usual density matrix
conditions such as normalization, positive semi-definiteness and
Hermiticity require that the diagonal elements $a,b,c,d$ are
non-negative and the equality $a+b+c+d=1$ holds. Moreover, the
anti-diagonal elements $w$ and $z$ satisfy
\begin{equation}
 |w|^2\leqslant ad,~ |z|^2\leqslant bc.
\end{equation}

If Bob is equipped to perform all kinds of unitary transformations
to the qubit at his possession, then the maximal average fidelity
achievable can be expressed as \cite{Horodecki1}
\begin{equation}
 F_{\rm max}^{(1)}=\frac{1}{2}+\frac{1}{6}N(\rho),
\end{equation}
where $N(\rho)={\rm tr}\sqrt{T^\dag T}$. Here $T$ is a $3\times 3$
positive matrix with elements $t_{nm}={\rm
tr}(\rho\sigma^n\otimes\sigma^m)$, and $\sigma^{1,2,3}$ are the
usual Pauli spin operators. $N(\rho)$ can be written explicitly as
$N(\rho)=\sum_{i=1}^3\sqrt{u_i}$, where $u_i$ $(i=1,2,3)$ are
eigenvalues of the matrix $T^\dag T$. Particularly, for the X state
expressed in Eq. (1), the eigenvalues of $T^\dag T$ can be obtained
straightforwardly as $u_{1,2}=4(|w|\pm|z|)^2$ and $u_3=(a+d-b-c)^2$,
thus we get $N(\rho)=2[|w|+|z|+|(|w|-|z|)|+|a+d-b-c|$.

If Bob can only perform the identity or the Pauli rotation
operations, then the maximal average fidelity for quantum
teleportation can be obtained explicitly as \cite{Yeo3,Horodecki2}
\begin{equation}
 F_{\rm max}^{(2)}=\frac{1}{3}+\frac{2}{3}\mathcal {F}(\rho),
\end{equation}
where $\mathcal {F}(\rho)={\rm max}\{\chi_0,\chi_1,\chi_2,\chi_3\}$
is the fully entangled fraction \cite{Bowen,Albeveio}, with the
notations $\chi_{0,3}=(a+d\pm 2|w|)/2$ and $\chi_{1,2}=(b+c\pm
2|z|)/2$. Clearly, $F_{\rm max}^{(2)}$ is in fact determined only by
the quantity $\chi_0$ or $\chi_1$. One can show now that for some
channel states that admit $F_{\rm max}^{(1)}>2/3$, the average
fidelity $F_{\rm max}^{(2)}$ may be equal to or smaller than $2/3$.
A representative example is the maximally entangled state
$|\varphi\rangle=(|00\rangle+|01\rangle+|10\rangle-|11\rangle)/2$
(this state can be generated by applying a Hadamard operation
\cite{Nielsen} to the second qubit of the Bell state
$|\Phi^+\rangle=(|00\rangle+|11\rangle)/\sqrt{2}$) which yields
$F_{\rm max}^{(1)}=1$ and $F_{\rm max}^{(2)}=2/3$.

For $2\times 2$ systems, the nonlocality of a quantum state can be
detected by the violation of the Bell-CHSH inequality proposed by
Clauser, Horne, Shimony and Holt \cite{Clauser}, which is given by
\begin{equation}
 |\langle B_{\rm CHSH}\rangle_\rho|\leqslant 2,
\end{equation}
where $\langle B_{\rm CHSH}\rangle_\rho={\rm tr}(\rho B_{\rm
CHSH})$, and $B_{\rm CHSH}$ is the Bell operator associated with the
quantum CHSH inequality. It has been demonstrated \cite{Horodecki3}
that $B_{\rm max}(\rho)={\rm max}|\langle B_{\rm CHSH}\rangle_\rho|$
is related to a quantity $M(\rho)$ via $B_{\rm
max}(\rho)=2\sqrt{M(\rho)}$, where $M(\rho)={\rm
max}_{i<j}(u_i+u_j)$, with $u_i$ $(i=1,2,3)$ being the eigenvalues
of the matrix $T^\dag T$. For the X state expressed in Eq. (1), we
can obtain
\begin{equation}
 M(\rho)={\rm max}\{8(|w|^2+|z|^2),4(|w|+|z|)^2+(a+d-b-c)^2\},
\end{equation}

Clearly, the inequality (5) is violated if and only if $M(\rho)>1$,
and the quantity $M(\rho)$ can also be used to measure the degree of
violation of the Bell nonlocality for a bipartite state.

We now begin our discussion about possible relations between
Bell-nonlocality violation and average fidelity $F_{\rm max}^{(2)}$
for the situation in which Alice performs her joint measurement
using the generalized Bell operators while Bob is only allowed to
perform the identity or the Pauli rotation operation. We will show
that if the Bell-CHSH inequality in Eq. (5) is violated, i.e.,
$B_{\rm max}(\rho)>2$ or $M(\rho)>1$, then all the X states will
yield $F_{\rm max}^{(2)}>2/3$. Since $F_{\rm max}^{(2)}$ is
determined only by $\chi_0$ or $\chi_1$, and it is easy to prove
that the two inequalities $\chi_0>1/2$ and $\chi_1>1/2$ cannot be
satisfied simultaneously. This is because if they are satisfied
simultaneously, then from the normalization of $\rho$ one can obtain
$\chi_0+\chi_1=(1+2|w|+2|z|)/2>1$, which gives rise to
$|w|+|z|>1/2$. Clearly, this is in contradiction with the fact that
$|w|+|z|\leqslant\sqrt{ad}+\sqrt{bc}\leqslant(a+d+b+c)/2=1/2$, which
can be derived directly from Eq. (2). Thus in the following we only
need to prove that the two inequalities $\chi_0<1/2$ and
$\chi_1<1/2$ cannot be satisfied simultaneously if $M(\rho)>1$.

The relative magnitude of $M(\rho)$ is determined by the maximum of
$M_1(\rho)=8(|w|^2+|z|^2)$ and $M_2(\rho)=4(|w|+|z|)^2+(a+d-b-c)^2$.
First, for the case of $M_1(\rho)\geqslant M_2(\rho)$, we have
$M(\rho)=8(|w|^2+|z|^2)$. If $\chi_0<1/2$ and $\chi_1<1/2$ are
satisfied simultaneously, then we get
\begin{equation}
 2|w|<b+c,~ 2|z|<a+d,
\end{equation}
where we have used the normalization condition $a+b+c+d=1$ in
deriving the above equations. Moreover, from Eq. (2) and the
positive semi-definiteness of the density matrix it is direct to
show that the following inequalities hold
\begin{equation}
 a+d\geqslant 2\sqrt{ad}\geqslant 2|w|,~ b+c\geqslant 2\sqrt{bc}\geqslant
 2|z|.
\end{equation}

By combination of Eqs. (7) and (8) one can obtain
\begin{equation}
 |w|<\frac{1}{4},~ |z|<\frac{1}{4},
\end{equation}
which yields $M(\rho)=8(|w|^2+|z|^2)<1$. Thus one see that for the
case of $M_1(\rho)\geqslant M_2(\rho)$, the Bell-CHSH inequality (5)
cannot be violated if $\chi_0<1/2$ and $\chi_1<1/2$.

Second, for the case of $M_1(\rho)<M_2(\rho)$ we have
$M(\rho)=4(|w|+|z|)^2+(a+d-b-c)^2$. Still one can prove that the
relations $\chi_0<1/2$ and $\chi_1<1/2$ cannot be satisfied
simultaneously. This is because we always have
\begin{equation}
 |w|+|z|<b+c,~ |w|+|z|<a+d,
\end{equation}
if $\chi_0<1/2$ and $\chi_1<1/2$, where the first inequality in Eq.
(10) can be obtained by combination of the first inequality in Eq.
(7) and the second inequality in Eq. (8), while the second
inequality in Eq. (10) can be obtained by combination of the second
inequality in Eq. (7) and the first inequality in Eq. (8). Because
the parameters appeared both in the left-hand side and the
right-hand side of the inequalities of Eq. (10) is non-negative, we
get
\begin{equation}
 (|w|+|z|)^2<(a+d)(b+c).
\end{equation}

On the other hand, violation of the Bell-CHSH inequality $|\langle
B_{\rm CHSN}\rangle_\rho|\leqslant 2$ for the X state $\rho$
requires $M(\rho)=4(|w|+|z|)^2+(a+d-b-c)^2>1$. It is easy to check
that this inequality can also be expressed equivalently as
$(|w|+|z|)^2>(a+d)(b+c)$, which is obviously contradicts the result
of Eq. (11). Thus by using apagogic reasoning we demonstrated again
that for the case of $M_1(\rho)<M_2(\rho)$, the Bell-CHSH inequality
still cannot be violated if $\chi_0<1/2$ and $\chi_1<1/2$.

Based on the above discussions, we came to the following proposition
about possible relations between Bell-nonlocality violation and
quantum teleportation.
\\

{\bf Proposition 1.} All the X states that violate the Bell-CHSH
inequality can be used for teleportation with average fidelity
$F_{\rm max}^{(2)}$ greater than the classical limiting value of
$2/3$.
\\

However, one should note that the inequality $B_{\rm max}(\rho)>2$
or $M(\rho)>1$ is only a sufficient condition for nonclassical
teleportation fidelity, because there are states which do not
violate the Bell-CHSH inequality, but still give rise to $F_{\rm
max}^{(2)}>2/3$. One such example is the Werner mixed state
\cite{Werner} described by the density matrix $\rho_{\rm
W}=p|\Psi^-\rangle\langle\Psi^-|+(1-p)\mathbb{I}_4/4$, where $p$ is
a real parameter ranges from 0 to 1,
$|\Psi^-\rangle=(|01\rangle-|10\rangle)/\sqrt{2}$ is the Bell
singlet state, and $\mathbb{I}_4$ denotes the $4\times 4$ identity
operator. For $\rho_{\rm W}$, from the above formulae one can obtain
$M(\rho)=2p^2$ and $F_{\rm max}^{(1)}=F_{\rm max}^{(2)}=(1+p)/2$.
Clearly, The Werner mixed state yields $F_{\rm max}^{(2)}>2/3$ for
$p>1/3$, while it violates the CHSH inequality only when
$p>1/\sqrt{2}$. Thus in the region of $p\in(1/3,1/\sqrt{2]}$ we have
the state $\rho_{\rm W}$ which are suitable for nonclassical quantum
teleportation but do not violate the Bell-CHSH inequality. Moreover,
it is straightforward to check that entanglement of the Werner state
$\rho_{\rm W}$ measured by the concurrence \cite{Wootters} is given
by $C={\rm max}\{0,(3p-1)/2\}$. This indicates that all the
entangled Werner states are useful for teleportation. But it should
be note that this is not the case for general forms of entangled X
states \cite{Hu1,Hu2}.

When considering relations between entanglement and Bell-nonlocality
violation, it has been shown by Verstraete and Wolf
\cite{Verstraete} that for states with a given concurrence $C$,
there exists an exact bound for $B_{\rm max}(\rho)$ :
$2\sqrt{2}C\leqslant B_{\rm max}(\rho)\leqslant 2\sqrt{1+C^2}$,
which shows clearly that the violation of the Bell-CHSH inequality
is guaranteed when $C>1/\sqrt{2}$. This relation also applies to the
X states considered here, namely, both the lower bound $2\sqrt{2}C$
and the upper bound $2\sqrt{1+C^2}$ for $B_{\rm max}(\rho)$ remain
unchanged for the X state. Thus we immediately came to the
conclusion that any X state with concurrence larger than
$1/\sqrt{2}$ is always useful for nonclassical teleportation, even
if Bob can only perform the identity (i.e., do nothing) or the Pauli
rotation operation. But it should be note that for the case of
$C<1/\sqrt{2}$, it is also possible to achieve nonclassical
teleportation fidelity (see, for example, the case of the Werner
state $\rho_{\rm W}$).

Now we turn to discuss possible relations between the average
fidelities $F_{\rm max}^{(1)}$ and $F_{\rm max}^{(2)}$. Since the
former represents the situation in which Bob is equipped to apply
any unitary transformation, while the latter corresponds to the
situation in which Bob can only perform the identity or the Pauli
rotation operation, we have $F_{\rm max}^{(1)}\geqslant F_{\rm
max}^{(2)}$ in general. What we concern in the following is the
extent to which $F_{\rm max}^{(2)}$ can be improved via Bob's
arbitrary transformations. For this purpose, we consider the
difference between $F_{\rm max}^{(1)}$ and $F_{\rm max}^{(2)}$,
i.e., $\delta F_{\rm max}=F_{\rm max}^{(1)}-F_{\rm max}^{(2)}$,
which can be derived straightforwardly as
\begin{equation}
 \delta F_{\rm max}=\frac{N(\rho)-4\mathcal {F}(\rho)+1}{6}.
\end{equation}

Since both the anti-diagonal elements $w$ and $z$ of $\rho$
contribute to $F_{\rm max}^{(1)}$ and $F_{\rm max}^{(2)}$ only in
the form of $|w|$ and $|z|$ (see section above), it suffice to
consider the special case of $\{w,z\}\in\mathbb{R}$, $w\geqslant 0$
and $z\geqslant 0$, and the conclusion obtained can be generalized
directly to the cases for general X states with negative or complex
anti-diagonal elements.

For $w\geqslant z$ and $\chi_0\geqslant\chi_1$, one can obtain
$N(\rho)-4\mathcal {F}(\rho)+1=|2(a+d)-1|-2(a+d)+1$, the relative
magnitude of which depends on the parameters $a$ and $d$ involved.
If $a+d\geqslant 1/2$, then we have $N(\rho)-4\mathcal
{F}(\rho)+1=0$ and $\delta F_{\rm max}=0$, i.e., during this
parameter region both $F_{\rm max}^{(1)}$ and $F_{\rm max}^{(2)}$
yield completely the same value. If $a+d<1/2$, however,
$N(\rho)-4\mathcal {F}(\rho)+1=2-4(a+d)$, which increases with
decreasing value of $a+d$. Because the assumed condition
$\chi_0\geqslant\chi_1$ requires $1+2z\leqslant 2(a+d)+2w$, and from
Eq. (2) one can obtain that $2w\leqslant 2\sqrt{ad}\leqslant a+d$,
thus we get $a+d\geqslant (1+2z)/3\geqslant1/3$. This, together with
the assumed condition $a+d<1/2$ gives rise to $N(\rho)-4\mathcal
{F}(\rho)+1\in(0,2/3]$ and $\delta F_{\rm max}\in(0,1/9]$.

For $w\geqslant z$ and $\chi_0<\chi_1$, it is straightforward to
check that $2(b+c)>1+2(w-z)\geqslant 1$, which gives rise to
$N(\rho)-4\mathcal {F}(\rho)+1=4(w-z)\leqslant 4w$. On the other
hand, from $\chi_0<\chi_1$ and Eq. (2) one can derive
$1+2z>2(a+d)+2w$ and $a+d\geqslant2\sqrt{ad}\geqslant 2w$, thus we
get $w<(1+z)/6\leqslant 1/6$, which gives rise to the upper bound of
$N(\rho)-4\mathcal {F}(\rho)+1$ as $2/3$. Moreover, the lower bound
of $N(\rho)-4\mathcal {F}(\rho)+1$ is $0$ because we have assumed
$w\geqslant z$. Thus by combining these results we get
$N(\rho)-4\mathcal {F}(\rho)+1\in[0,2/3)$ and $\delta F_{\rm
max}\in[0,1/9)$.

From the above analysis one can see that during the parameter region
$w\geqslant z$, the difference between $F_{\rm max}^{(1)}$ and
$F_{\rm max}^{(2)}$ ranges from $0$ to $1/9$, i.e., $\delta F_{\rm
max}\in[0,1/9]$. The maximal difference $\delta F_{\rm max}=1/9$
occurs only when the involved matrix elements satisfying $a+d=1/3$,
$b+c=2/3$, $w=1/6$, and $z=0$, which corresponds to $F_{\rm
max}^{(1)}=2/3$ and $F_{\rm max}^{(2)}=5/9$. Since $F_{\rm
max}^{(2)}>2/3~(\neq 5/9)$ is guaranteed if $B_{\rm max}(\rho)>2$,
we can also conclude that for X states which violate the Bell-CHSH
inequality, the difference between $F_{\rm max}^{(1)}$ and $F_{\rm
max}^{(2)}$ must be smaller than $1/9$.

Moreover, for the case of $w<z$, one can still obtain $\delta F_{\rm
max}\in[0,1/9]$ after a similar analysis as performed for that of
$w\geqslant z$, with however the maximal difference $\delta F_{\rm
max}=1/9$ occurs when $a+d=2/3$, $b+c=1/3$, $w=0$, and $z=1/6$. Thus
in light of the above results, we can draw the following
proposition.
\\

{\bf Proposition 2.} For all the X states, the maximal difference
between $F_{\rm max}^{(1)}$ and $F_{\rm max}^{(2)}$ is $1/9$.
\\

This Proposition represents the extent to which the average fidelity
can be improved by Bob's arbitrary transformations. Particularly,
for the case of $F_{\rm max}^{(2)}\in(5/9,2/3]$, if Bob is equipped
to perform some unitary transformations other than that of the
identity or the Pauli rotation operation, then the average fidelity
can be enhanced to over its classical limiting value of $2/3$.

Before ending this paper, we would also like to see fractions of
different types of X states over the ensemble of the X states. Since
all the X states which violate the Bell-CHSH inequality are
entangled and useful for teleportation, while there also exist X
states which satisfy the Bell-CHSH inequality but still can be used
for teleportation, we can draw the conclusion that $P_{\rm E}>P_{\rm
T}>P_{\rm B}$, where $P_{\rm E}$, $P_{\rm T}$ and $P_{\rm B}$
denote, respectively, fraction of the entangled X states, fraction
of the X states that are useful for nonclassical teleportation and
fraction of the X states that violate the Bell-CHSH inequality.

\section{Summary}
In summary, we've studied possible relations between entanglement,
Bell-CHSH inequality violation and quantum teleportation for the X
states. As a generalization of the work \cite{Horodecki1} in which
the authors proved that any two spin-1/2 state which violates the
Bell-CHSH inequality is useful for teleportation if Bob is equipped
to perform all types of the unitary transformations, here we further
demonstrated that for the X states, nonclassical teleportation is
also guaranteed by violation of the Bell-CHSH inequality even if Bob
can only perform the identity or the Pauli rotation operations.
Since the X states occur in many contexts \cite{Nielsen,Werner} and
experimental realization of the Pauli rotation is comparatively
simple (see \cite{Nielsen} and references therein), we hope our
results will be relevant to the practical teleportation process.
Moreover, we also compared the maximal average fidelities $F_{\rm
max}^{(1)}$ and $F_{\rm max}^{(2)}$, which associate to the
situations in which Bob is allowed to perform any unitary
transformation and Bob can only perform the identity or the Pauli
rotation operations on his qubit, respectively. Our results revealed
that the difference between them ranges from 0 to $1/9$, where the
upper bound $1/9$ represents the greatest extent to which the
average fidelity can be improved via Bob's arbitrary
transformations.
\\

\begin{center}
\textbf{ACKNOWLEDGMENTS}
\end{center}

This work was supported by NSF of Shaanxi Province under grant Nos.
2010JM1011 and 2009JQ8006, the Specialized Research Program of
Education Department of Shaanxi Provincial Government under grant
Nos. 2010JK843 and 2010JK828, and the Youth Foundation of XUPT under
Grant No. ZL2010-32.

\newcommand{\PRL}{Phys. Rev. Lett. }
\newcommand{\PRA}{Phys. Rev. A }
\newcommand{\PLA}{Phys. Lett. A }
%

%

\end{document}